\documentclass[runningheads]{llncs}
\usepackage{graphicx}
\usepackage{array}
\usepackage{booktabs}
\usepackage{diagbox}
\usepackage[table,xcdraw]{xcolor}
\usepackage{tabularx}
\usepackage{subfig}
\usepackage{wrapfig}
\usepackage[title]{appendix}
\begin{document}
\newcolumntype{L}[1]{>{\raggedright\arraybackslash}p{#1}}
\newcolumntype{C}[1]{>{\centering\arraybackslash}p{#1}}
\newcolumntype{R}[1]{>{\raggedleft\arraybackslash}p{#1}}
\title{Towards a Context-Aware Security and Privacy as a Service in the Internet of Things}
\titlerunning{Towards a Context-Aware Security and Privacy as a Service in the IoT}
%
\author{Tidiane Sylla\inst{1,3}\orcidID{0000-0002-3781-6973} \and
Mohamed Aymen Chalouf\inst{2}\orcidID{0000-0002-4653-7837} \and
Francine Krief\inst{1}\orcidID{0000-0003-4702-9305}
\and Karim Samaké\inst{3}}
\authorrunning{Tidiane Sylla, Mohamed Aymen Chalouf, Francine Krief, Karim Samaké}
%
\institute{Univ. Bordeaux, Bordeaux INP, CNRS, LaBRI, UMR 5800, 33400 Talence, France \and
Univ. Rennes 1, CNRS, IRISA Lab, UMR 6074, 22300 Lannion, France\and
Univ. Sciences Techniques and Technologies Bamako, Bamako, Mali}

\maketitle              
\begin{abstract}
Smart city is one of the most known Internet of Things (IoT) applications. The smart city services improve user’s daily lives. However, security and privacy issues are slowing down their adoption. In addition, the characteristics of IoT devices, applications and users make security implementation of the considered applications a challenging task. To address these issues, we present, in this paper, a new context-aware security and privacy architecture for the IoT. Thanks to the ``as a service" approach, this new architecture will be user-centric. It will also support known context-aware security issues: dynamicity, flexibility. In addition, it will address mobility, customization of security and privacy services, and support for generic IoT applications, particularly for smart city. To do so, a knowledge plane allowing effective management of context-awareness is proposed. A security and privacy plane allowing better implementation of context-aware security and privacy mechanisms is also proposed. This will be done through dynamic composition of context-based micro services. The role of the different components of these two planes are also described.

\vspace*{-0.3cm}
\end{abstract}
\section{Introduction} 
Internet of Things (IoT) applications enable advanced and intelligent services that make users everyday life easier. In this work, we are interested in the smart city field. It is a topical field and it includes a number of interesting IoT applications such as e-health, smart home, vehicular networks, etc. The implementation of smart city IoT applications and devices may involve risks related to the users's security and privacy (disclosure, espionage, theft, etc.). These problems have been addressed in a large number of works \cite{alagar_context-based_2018,de_matos_providing_2018,neisse_dynamic_2015}.\\
\indent However, these solutions do not consider user’s characteristics, such as privacy preferences, mobility, usability, etc. To overcome these problems, the emphasis should be on a user-centric approach. Due to its importance and relevance for IoT and other digital services, the European Telecommunications Standards Institute (ETSI) has adopted several standards\cite{aubonnet_user_2019}. Indeed, it allows users to play a central role in security and privacy. Thus, implementing security and privacy mechanisms according to some relevant information about users (e.g. contexts) and without their explicit intervention become necessary.\\
\indent Furthermore, the security and privacy mechanisms specified in many research works are proposed or implemented to address specific security threat models to which the targeted system is exposed. Since the situation of a considered user could change due to many factors (e.g. mobility), the threat models will also change. Therefore, to ensure optimal security and address the detected vulnerabilities properly, the implementation of several security mechanisms is necessary according to different user situations.\\
\indent Context-aware security and privacy is an effective way to implement user-centric security and privacy. It will allow to manage the threat models related to the users’ frequent context changes. This is done by dynamically deploying security and privacy mechanisms that respond to the threat model characterizing user's current context without his intervention.\\
\indent In this regard, different proposals have been introduced. However, to the best of our knowledge, none of these works propose a solution that meets the requirements : secure context-awareness management, privacy, authentication, access control and communication security. In addition, to meet next generation networks architecture requirements, security and privacy of the IoT could be based on the ``\textbf{as a service}" approach. This allows it to provide flexibility, dynamicity, scalability, and better support for user mobility and heterogeneity \cite{aubonnet_controlled_2016}. \\
\indent That is why our work goes beyond existing works, by proposing a \textbf{Context-Aware Security and Privacy as a Service} (\textbf{CASPaaS}) based architecture. The main innovations of our work are the introduction of a knowledge plane, responsible for managing context-awareness through Machine Learning (ML) and Quality of Context (QoC), and a security and privacy plane, responsible for implementing mechanisms through the dynamic composition of context-based micro services.\\
\indent The rest of the paper is organized as follows. Section 2 presents and compares related works. Section 3 describes our contribution. Finally, section 4 concludes the paper and presents our further works.
\section{Related works}
Context-aware security and privacy in the IoT has been the subject of several studies. In this section, we compare the different proposed solutions and point out the remaining challenges.
\vspace*{-1mm}
\subsection{Proposed solutions}
A context-aware security and privacy solution in smart city IoT applications has been proposed in \cite{neisse_dynamic_2015}. This solution implements context-based security policy management. It uses a combination of several contextual parameters (time, location, network, speed) for context perception. It allows the user to define some preferences (e.g. access control). The use of a combination of several contextual parameters can help to determine the context with greater precision. However, this paper has only focused on the implementation of policy-based security. It does not support the security of contextual information management. Thus, this mechanism is vulnerable to attacks of identity theft and fake location.\\ 
\indent The solution described in \cite{ramos_managing_2015} also implements context-aware security and privacy. Unlike the solution proposed in \cite{neisse_dynamic_2015}, the proposed context-awareness management system implements context information security. Nevertheless, Quality of Context (QoC) is not taken into account in these solutions. Thus, the contexts determined by these solutions can be subject to conflicts.\\
\indent Context-aware privacy is complementary to context-aware security in the IoT. Therefore, in \cite{neisse_dynamic_2015}, the authors described a privacy mechanism based on pseudo-anonymization and delayed message delivery. Delayed message delivery can prevent user tracking (e.g. in geolocation). In \cite{ramos_managing_2015}, the authors presented a privacy system based on the anonymization of user's data. However, pseudo-anonymization and anonymization are vulnerable to inference attacks on user data. In \cite{de_matos_providing_2018}, a context-aware security module offering privacy is described. However, the authors did not provide details on the technique used in this module.\\
\indent In \cite{ashibani_context-aware_2017}, the authors focused on context-aware authentication. The proposed mechanism uses a combination of username/password as an authentication factor, making it vulnerable to passwords attacks. In addition, the authors of \cite{de_matos_providing_2018} and \cite{neisse_dynamic_2015} addressed authentication and access control. However, the context-aware security module proposed in \cite{de_matos_providing_2018} does not specifically define how authentication and access control are sensitive to the context.\\
\indent In \cite{ramos_managing_2015}, the authors proposed an access control mechanism based on contextual access tokens. However, this mechanism does not enable user to dynamically define authorizations. Moreover, it does not have the needed flexibility to support the aforementioned features. In addition, the authorization management system is centralized, which can cause a single point of failure. \\
\indent Context-aware communication security allows secure communications irrespective of whether the underlying networks are secured or not. However, none of the studied works proposed a mechanism for  communication security.
\vspace*{-2mm}
\subsection{Positioning}
The above-described works propose context-aware security solutions in the IoT. Table~\ref{tab1} summarizes the comparison between these works. On the one hand, the support of the proposed contextual security and privacy mechanisms are mostly incomplete for the IoT. On the other hand, beyond these challenges, these works addressed the issues of context-aware security and privacy in a specific application field. In IoT, each user can have several devices and applications. Thus, proposing an architecture that allows to meet the requirements identified independently of smart city IoT applications and devices becomes necessary.\\
\begin{table}[!htbp]
\caption{Comparison of work that has proposed context-aware security and privacy solutions in the IoT}\label{tab1}
\resizebox{\textwidth}{!}{%
\begin{tabular}{@{}L{1.525cm}C{1.525cm}C{1.525cm}C{1.525cm}C{1.525cm}C{1.525cm}C{1.525cm}C{1.525cm}@{}}
\toprule
\diagbox[innerwidth=1.8cm, innerleftsep=0pt, innerrightsep=0cm]{Works}{C.A.S} & C.A Authentication&C.A Authorization&C.A Commu. security&C.A Privacy&Context mgmt. security & As a Service & ITU-T ref. arch. integration\\ \midrule
\multicolumn{1}{|l|}{\cite{neisse_dynamic_2015}} & \multicolumn{1}{c|}{\cellcolor[HTML]{f7c965}Mentioned} & \multicolumn{1}{c|}{\cellcolor[HTML]{f7c965}Mentioned} & \multicolumn{1}{c|}{\cellcolor[HTML]{FE996B}No} & \multicolumn{1}{c|}{\cellcolor[HTML]{3fc690}Yes} & \multicolumn{1}{c|}{\cellcolor[HTML]{FE996B}No} & \multicolumn{1}{c|}{\cellcolor[HTML]{FE996B}No} & \multicolumn{1}{c|}{\cellcolor[HTML]{FE996B}No} \\ \midrule
\multicolumn{1}{|l|}{\cite{ramos_managing_2015}} & \multicolumn{1}{c|}{\cellcolor[HTML]{FE996B}No} & \multicolumn{1}{c|}{\cellcolor[HTML]{3fc690}Yes} & \multicolumn{1}{c|}{\cellcolor[HTML]{FE996B}No} & \multicolumn{1}{c|}{\cellcolor[HTML]{f7c965}Mentioned} & \multicolumn{1}{c|}{\cellcolor[HTML]{3fc690}Yes} & \multicolumn{1}{c|}{\cellcolor[HTML]{FE996B}No} & \multicolumn{1}{c|}{\cellcolor[HTML]{FE996B}No} \\ \midrule
\multicolumn{1}{|l|}{\cite{ashibani_context-aware_2017}} & \multicolumn{1}{c|}{\cellcolor[HTML]{3fc690}Yes} & \multicolumn{1}{c|}{\cellcolor[HTML]{f7c965}Mentioned} & \multicolumn{1}{c|}{\cellcolor[HTML]{FE996B}No} & \multicolumn{1}{c|}{\cellcolor[HTML]{FE996B}No} & \multicolumn{1}{c|}{\cellcolor[HTML]{FE996B}No} & \multicolumn{1}{c|}{\cellcolor[HTML]{FE996B}No} & \multicolumn{1}{c|}{\cellcolor[HTML]{FE996B}No} \\ \midrule
\multicolumn{1}{|l|}{\cite{de_matos_providing_2018}} & \multicolumn{1}{c|}{\cellcolor[HTML]{f7c965}Mentioned} & \multicolumn{1}{c|}{\cellcolor[HTML]{f7c965}Mentioned} & \multicolumn{1}{c|}{\cellcolor[HTML]{FE996B}No} & \multicolumn{1}{c|}{\cellcolor[HTML]{f7c965}Mentioned} & \multicolumn{1}{c|}{\cellcolor[HTML]{FE996B}No} & \multicolumn{1}{c|}{\cellcolor[HTML]{FE996B}No} & \multicolumn{1}{c|}{\cellcolor[HTML]{FE996B}No} \\ \midrule
\multicolumn{1}{|l|}{\cite{alagar_context-based_2018}} & \multicolumn{1}{c|}{\cellcolor[HTML]{FE996B}No} & \multicolumn{1}{c|}{\cellcolor[HTML]{3fc690}Yes} & \multicolumn{1}{c|}{\cellcolor[HTML]{FE996B}No} & \multicolumn{1}{c|}{\cellcolor[HTML]{3fc690}Yes} & \multicolumn{1}{c|}{\cellcolor[HTML]{FE996B}No} & \multicolumn{1}{c|}{\cellcolor[HTML]{FE996B}No} & \multicolumn{1}{c|}{\cellcolor[HTML]{FE996B}No} \\ 
\midrule
\multicolumn{1}{|l|}{\cite{barhamgi_user-centric_2018}} & \multicolumn{1}{c|}{\cellcolor[HTML]{FE996B}No} & \multicolumn{1}{c|}{\cellcolor[HTML]{FE996B}No} & \multicolumn{1}{c|}{\cellcolor[HTML]{FE996B}No} & \multicolumn{1}{c|}{\cellcolor[HTML]{3fc690}Yes} & \multicolumn{1}{c|}{\cellcolor[HTML]{FE996B}No} & \multicolumn{1}{c|}{\cellcolor[HTML]{FE996B}No} & \multicolumn{1}{c|}{\cellcolor[HTML]{FE996B}No} \\ \midrule
\multicolumn{1}{|l|}{Proposition} & 
\multicolumn{1}{c|}{\cellcolor[HTML]{3fc690}Yes} & \multicolumn{1}{c|}{\cellcolor[HTML]{3fc690}Yes} & \multicolumn{1}{c|}{\cellcolor[HTML]{3fc690}Yes} & \multicolumn{1}{c|}{\cellcolor[HTML]{3fc690}Yes} & \multicolumn{1}{c|}{\cellcolor[HTML]{3fc690}Yes} & \multicolumn{1}{c|}{\cellcolor[HTML]{3fc690}Yes} & \multicolumn{1}{c|}{\cellcolor[HTML]{3fc690}Yes} \\ \bottomrule
\end{tabular}%
}
\vspace*{-0.59cm}
\end{table}
\indent Furthermore, the need to move towards Software Oriented Architecture (SOA) in the IoT is growing. On one hand, this is mainly due to the fact that SOA enables component-based model. SOA allows designing a system into functional parts \cite{aubonnet_controlled_2016}. On second hand, next generation networks are essentially software defined. The architecture proposed supports context-aware security requirements. Moreover, it addresses challenges such as dynamicity, flexibility, mobility, customization, and support for generic IoT applications through secure API, particularly for smart city. 
\section{Context-Aware Security and Privacy as a Service based architecture}
In this section, we give a detailed description of our contribution. We also highlight main challenges related to the architecture implementation.
\subsection{Overview}
An effective context-aware security and privacy needs a separation between the context-awareness management and the implementation of security and privacy mechanisms. Indeed, the separation of the intelligence (i.e, context-awareness) and the enforcement of security/privacy decisions enables more modularity and flexibility. Thus, these features enable more dynamicity and adaptability in offering security and privacy to users. Therefore, the proposed architecture is divided into two planes: \textbf{Knowledge Plane} (\textbf{KP}) and \textbf{Security and Privacy Plane} (\textbf{SPP}). These planes will integrate ITU-T IoT reference architecture to provide context-awareness and adaptive security and privacy (See Appendix A).\\
\indent Thanks to the ``as a service" approach, the architecture can be integrated into new service-oriented networks. Its addresses several challenges in securing the IoT (Section 2.2). Therefore, the modules composing the different planes are designed according to Virtual Network Function (VNF) requirements presented in  \cite{boubendir_flexibility_2018}. As a result, security and privacy for IoT applications will be dynamic, flexible, customizable and user-centric. \\
\begin{figure}[!ht]
    \centering
    \subfloat[Knowledge Plane\label{subfig-1:casaas}]{%
    \includegraphics[width=0.485\textwidth]{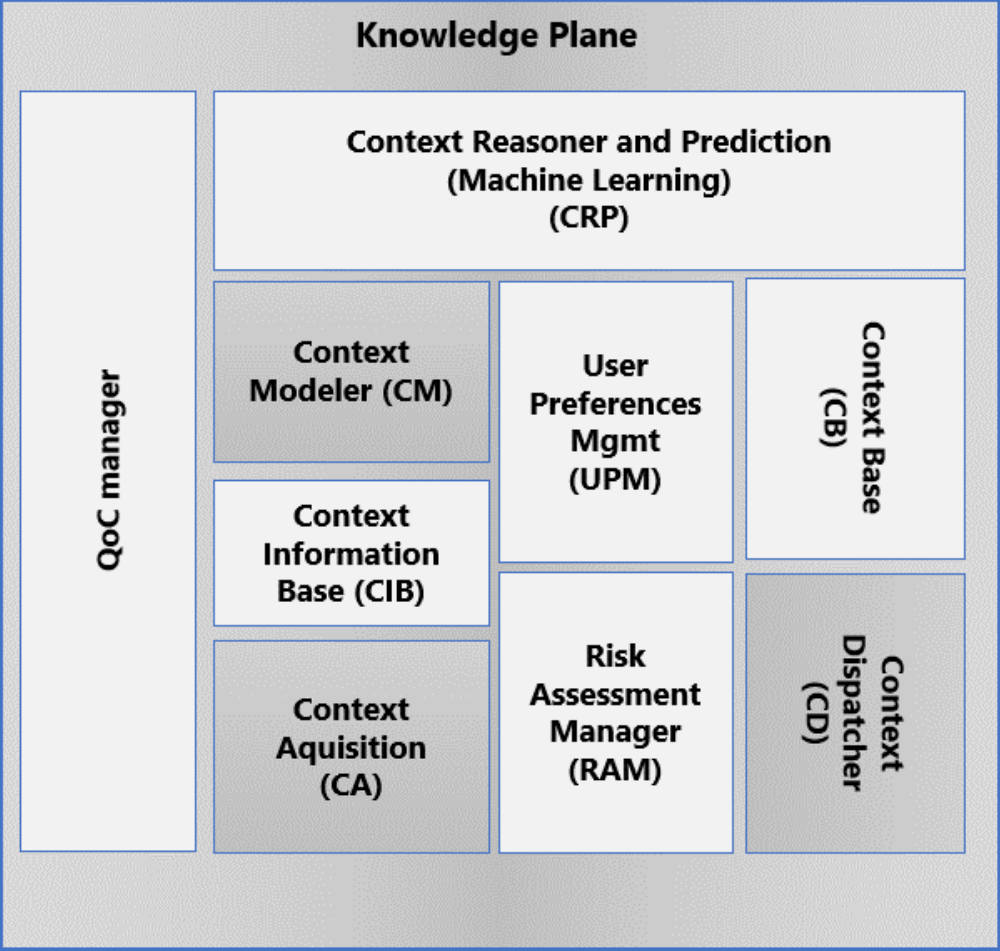}
    }
    \hfill
    \subfloat[Security and Privacy Plane\label{subfig-2:casaas}]{%
    \includegraphics[width=0.49\textwidth]{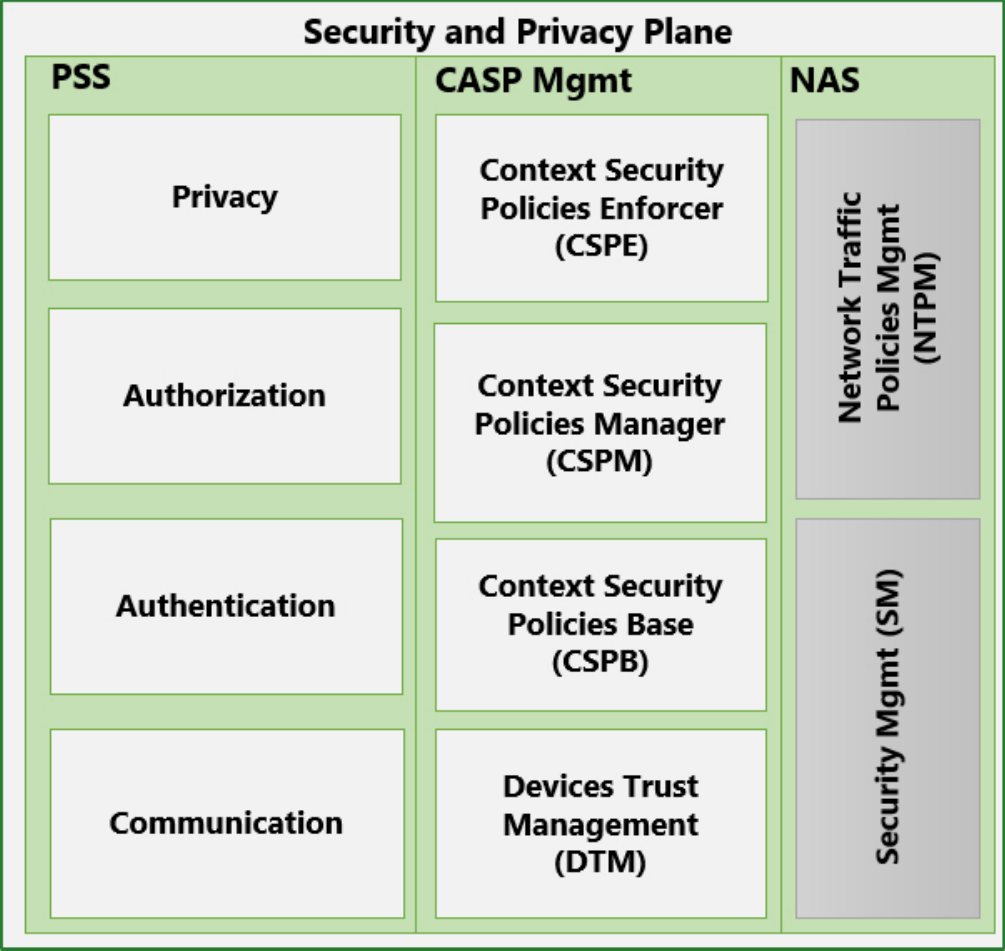}
    }
    \caption{Context-Aware Security as a Service Architecture} \label{fig2:casaas}
    \vspace*{-0.6cm}
\end{figure}
The walking through example of our architecture operation will be the following. Bob is a diabetic patient living in a smart home. He is equipped with a smart watch, which continuously monitors his glucose level and daily activities. The hospital’s smart healthcare system collects and processes Bob’s health information in order to provide him with better healthcare and feeding.
\subsection{Knowledge Plane}
\begin{figure}
	\centering
\includegraphics[width=0.8\textwidth]{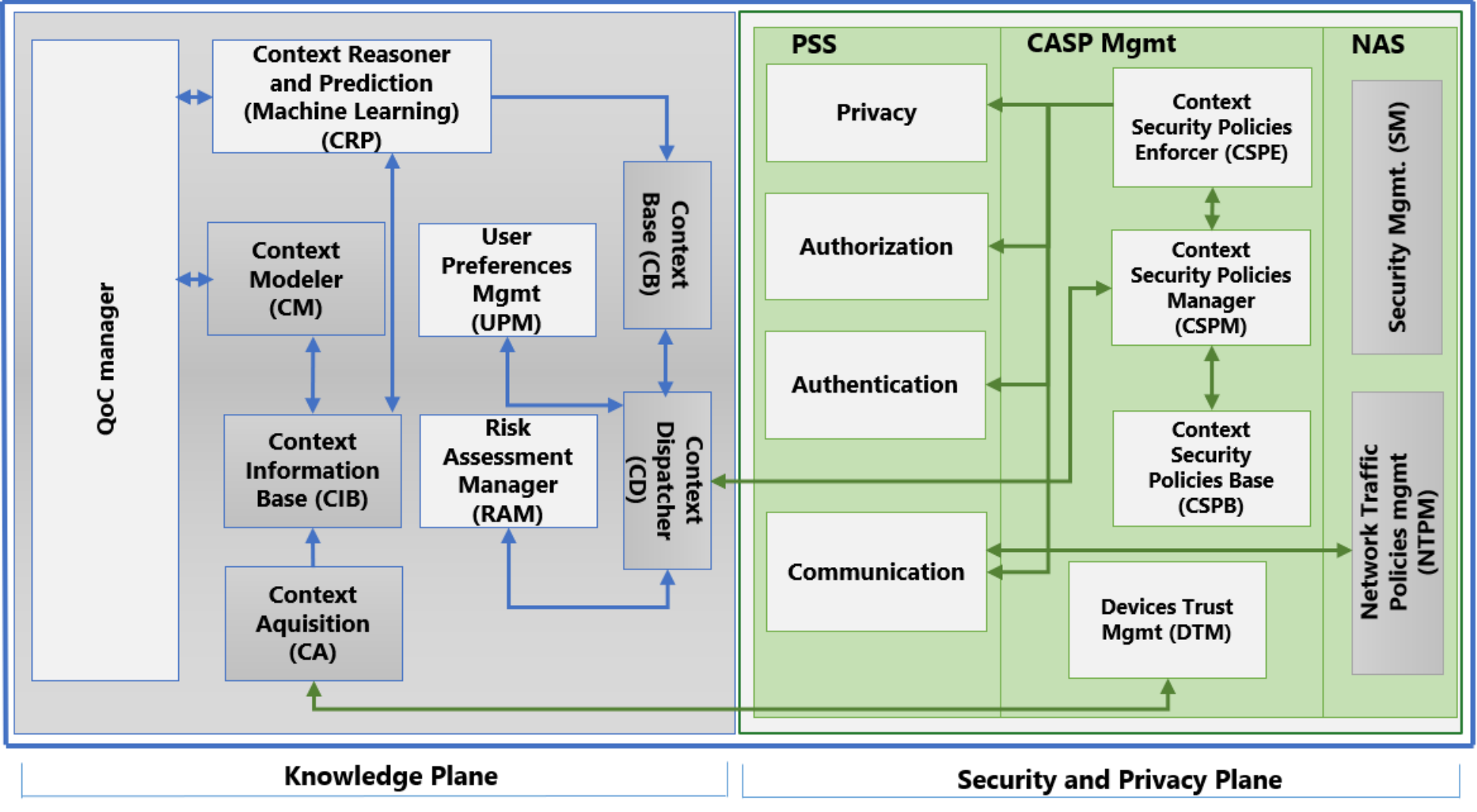}
\caption{CASPaaS modules and their interactions} \label{fig5}
\vspace*{-0.6cm}
\end{figure}

\indent The Knowledge Plane (KP) (Figure~\ref{fig2:casaas}a) aims to provide specific and relevant context and related information (e.g. risk level and preferences) to the SPP. Based on this, the SPP will implement appropriate security and privacy mechanisms. It is composed of modules necessary for the management of context life cycle, i.e., context acquisition, modelling, reasoning and dissemination \cite{perera_context_2014}. \\
\indent The first stage of context life cycle is context acquisition. The \textbf{Context Acquisition} (\textbf{CA}) module receives context information from trusted context sources (see Section 3.3). We refer by context source any device in the user's environment collecting context information. The CA module pre-processes (for example a raw GPS sensor data must be put in a format that represents geographical location) and stores context information, also called low-level context, in the \textbf{Context Information Base} (\textbf{CIB}). For example, Bob leaves his house and is walking in the street. In this case, following low-level contexts are sent to the service: date and time, Bob’s location, Bob’s network and motion.\\
\indent The next step in the context processing is context modelling. This is done by the \textbf{Context Modelling} (\textbf{CM}) module. Indeed, it represents the context in terms of context attributes, characteristics and \textbf{Quality of Context} (\textbf{QoC}) attributes. Then, the representation obtained is organized according to the chosen model. Different context models exist: the key-value model, ontology-based model, hybrid model, etc. \cite{perera_context_2014}. The choice of a model depends on the its ability to meet the requirements of the context modelling and the target application domain. In the considered example, a key-value model is well adapted to the situation because of its simplicity and flexibility in modelling such a situation. These operations are performed in collaboration with the QoC module.\\
\indent The QoC module aims to resolve conflicts in context determination. It is characterized by a set of parameters. First, the module computes QoC parameters (timeliness, reliability, completeness, importance) to measure the quality of the low-level context received. Then, the results of these measurements will be interpreted to determine the existence of conflicts. Depending on the type of detected conflict, it applies a set of policies to provide a context with a better-quality. For example, user's location sensing policy can be based on the up-to-dateness.\\
\indent After context modelling, the next stage in context management is the context reasoning. Context reasoning is the process of deduction high-level context from several low-level context information. The output of the CM is used by the \textbf{Context Reasoning and Prediction} (\textbf{CRP}) module to determine the high-level context. Indeed, it infers on the low-level context information provided by the CM using a context reasoning technique. In Bob’s case, the resulting high-level context will be: “\textit{user is walking near the home}”. There are several context reasoning techniques, including ontology-based, machine learning, fuzzy logic, etc. In our architecture, supervised learning technique will be used by the CRP module, because of its good accuracy. The determined high-level context is first validated by the QoC module. Then, the resulting high-level context is stored in the \textbf{Context Base} (\textbf{CB}).\\
\indent Finally, the last stage of the context management is the dissemination of high-level contexts. Before context dissemination, the KP will assess the risk level and user’s preferences associated with the context. These operations are performed by the \textbf{Risk Assessment Manager} (\textbf{RAM}) and the \textbf{User Preferences Management} (\textbf{UPM}) modules. Then, the context, risk level and user’ preferences will be straightforward distributed to the SPP for making contextual security decision. This context distribution is done by \textbf{Context Dispatcher} (\textbf{CD}). The main context consumer in the SPP is the \textbf{Context Security Policies Manager} (\textbf{CSPM})(Section 3.3). The dissemination of context and related information to the CSPM is done through a publish/subscribe mechanism.\\
\indent The RAM compute the risk level of a given context based on the threat model associated to that context. In the considered example, Bob is at a public garden with his friends. Bob’s devices (smartphone, smartwatch) are connected to the public garden Wifi network. After the CD receives Bob’s new context, it sends it to the RAM for risk assessment. The RAM assesses the given context risk based on its threat model (unsecure network, eavesdropping, etc.), so high security risk in Bob’s case. Next, the RAM returns to the CD Bob’s context with the assessed risk level. When the CD receives the context risk level, it gets the corresponding user’ preferences from the UPM and sends them to the CSPM. The SPP can use this new context and deploy appropriate security and privacy mechanisms. Thus, the KP provides the necessary intelligence to the SPP. Figure~\ref{fig5} illustrates interactions between the architecture components.
\subsection{Security and Privacy Plane}
The Security and Privacy Plane (Figure~\ref{fig2:casaas}b) addresses the identified context-aware security and privacy functional requirements. It is divided into three functional components: \textbf{Privacy and Security Services} (\textbf{PSS}), \textbf{Context-Aware Security and Privacy Management} (\textbf{CASP Mgmt}) and \textbf{Network and Architecture Security} (\textbf{NAS}).\\ 
\indent The PSS and CASP Mgmt components constitute the core of the SPP. Indeed, the CASP Mgmt includes the modules in charge of contextual security policies management and security of context-awareness management. The PSS is composed of modules responsible for the enforcement of contextual security and privacy decisions taken by the CASP Mgmt. Finally, the NAS includes modules providing architecture and network security.\\
\indent To provide secure context-awareness management, the architecture should be able to gather secured contexts from trusted user IoT devices. The \textbf{Device Trust Management} (\textbf{DTM}) is in charge of the management of contexts security and the trustworthiness of context sources. First, context sources will send encrypted context information to the DTM. A lightweight public key cryptography for IoT devices will be used to this end. Second, user’s devices trustworthiness should be established for each exchange. This will be based on devices reputation. Device reputation will be assessed by computing the trustworthiness of context it has sent. Third, the user should be able to manage his devices ownership. The \textbf{Blockchain} can be leveraged to achieve these goals. This choice is motivated by its following features. Firstly, Blockchain-based decentralized PKI (Public Key Infrastructure) is well suited for IoT \cite{axon2016pb,won2018decentralized}. Secondly, \textbf{smart contracts} features such as automated execution, transfer of property can help in automatic reputation assessment \cite{skulimowski_towards_2018}. It can also allow the user to control his device's ownership.\\
\indent A core element of the context-aware security and privacy is the management of contextual security policies. Thus, the \textbf{Context Security Policy Manager} (\textbf{CSPM}) is in charge of selecting the contextual security policy corresponding to a given context and related information (risk level, preferences). To do so, when the CSPM receives a context and related information, it gets the corresponding policy from the \textbf{Context Security Policies Base} (\textbf{CSPB}) and sends it to the \textbf{Context Security Policies Enforcer} (\textbf{CSPE}). \\
\indent The contextual security policy describes the security and privacy mechanisms to be deployed in a specific context. The role of the CSPE module is to use the security policy provided by the CSPM to order the enforcement of appropriate security and privacy mechanisms. This enforcement will be done by the modules of the PSS component. When a \textbf{Contextual Security Policy} (\textbf{CSP}) has to be enforced, the CSPE will orchestrate the composition of micro services corresponding to the appropriate modules of the PSS component. For example, the CSP can dictate the enforcement of the following mechanisms: \textit{two factor authentication, renew devices authentication keys, and secure communication}. \\
\indent After the contextual security policy decision processing, the selected policy must be enforced by context-aware security and privacy mechanisms. Thus, the \textbf{Privacy}, \textbf{Authentication}, \textbf{Authorization} and \textbf{Communication} modules are responsible of implementing these mechanisms. Besides, APIs will be provided to ensure the genericity of the solution and its independence from the IoT applications. This will enable the developers to export the security task of their applications by calling the provided APIs. \\
\indent The \textbf{Privacy} module will act as a privacy assistant. It will be able to continuously analyze the data coming from user’s devices. Depending on the context, it informs user if there is a proven risk to his privacy. It  also implements the rules provided by the CSPE.\\
\indent The \textbf{Authentication} module is in charge of users and IoT devices. Thus, according to the rules provided by the CSPE, a type of authentication is proposed to the user (e.g. one factor, double factor). For a device, depending on the context, the session key can be renewed. \\
\indent The \textbf{Authorization} module will manage resources access control. To this end, \textbf{Blockchain} can enables to define and manage the authorizations of an entity in a distributed way. This can be done according to the operation of an IoT application and based on user-centric approach. Indeed, an entity's authorizations must be represented in the form of tokens. Then it is entered in a smart contract registered in the Blockchain. Through the UPM, the user should be able to modify or revoke an authorization at any time. In all cases, the authorization is dynamically updated and implemented by the module.\\
\indent Communication security is needed in some contexts, especially for unsecured networks. Thus, the \textbf{Communication} module role is, according to a context, to secure communications between devices and applications by enforcing the associated CSP. This can be done by implementing message security (payload) of the application layer. Indeed, the effectiveness of message security in providing secure communications to IoT devices over unsecured networks is proven\cite{claeys_securing_2017}. Let’s suppose that the hospital healthcare system needs to pull Bob’s glucose level. Bob’s context is \textit{at the public garden}. For this context, the CSPM provides a CSP specifying \textit{secure communication} and privacy to the CSPE. The result of that is the establishment of secure communications between Bob’s smartwatch and the hospital’s healthcare system prior to any data transmission. After secure communication’s setup, Bob’s glucose level is anonymized/obscured.
\\\indent Finally, the architecture should be virtualized and deployed as a service. To this end, it must be secured in order to prevent possible attacks (e.g. denial of service). The role of \textbf{Security Management} (\textbf{SM}) is to ensure the security of the entire architecture. It implements a firewall and deep packet inspection for mitigating attacks against availability. It also addresses the user's mobility and devices heterogeneity. To do so, CSP rules will be sent to devices by leveraging SDN (Software Defined Network) capabilities.\\
\indent The \textbf{Network Traffic Policy Management} (\textbf{NTPM}) module is responsible for transmitting rules to devices. It dictates to the SDN controller the traffic paths based on the results provided by the SM in case of an attack. The devices will then act as SDN agents, capable of applying and redirecting traffic at the request of a SDN controller. The SDN controller will receive commands from the architecture's mechanism implementation components. Please see Appendix-B for an illustration of our architecture possible deployment in a network with an edge computing infrastructure.
\section{Conclusion and future work}
Context-aware security and privacy makes it possible to support the smart city IoT applications user’s situations changes. We have identified important points that should be considered : intelligence, security services and privacy, dynamicity, flexibility, scalability, mobility, genericity, scalability. \\
\indent In this sense, different solutions have been proposed. However, none of them have addressed the identified requirements. Hence, in this paper, these different approaches are described and compared, and a new architecture is proposed. This architecture, unlike the previous proposals, is designed based on ``as a service" approach. It is composed of two planes. Essentially, a Knowledge Plane, using QoC, Machine Learning and Risk management and improving context-awareness, is proposed. Besides, the devices trust management within Security and Privacy Plane is proposed.\\
\indent Future work will focus on the following points. The first objective is the implementation of the Device Trust Management module announced in section 3.3. Then, we will implement the authorization management module based on the Blockchain through a smart contract and contextual access tokens. This implementation will be based on the Hyperledger Fabric which is a Blockchain framework allowing the creation of smart contracts using Java language. Finally, we will perform a simulation of the architecture deployment in a 5G network and its performance evaluation will be performed.
%
%
%
%
\bibliographystyle{splncs04}
\bibliography{references}
\newpage
\begin{subappendices}
\renewcommand{\thesection}{\Alph{section}}%
\section{ITU-T Reference Architecture Integration}
The ITU-T IoT reference architecture integrates a transversal layer to ensure security across the different layers of the reference architecture\footnote{ITU-T Recommendation Y.4000/Y.2060, 2012}. Our proposed architecture aims to integrate this layer as a specific security capability in order to provide a context-aware security as a service for IoT. It also aims to integrate a knowledge plane in the ITU-T IoT reference architecture to enable context-awareness features for the management layer. Thus, our work will allow the ITU-T IoT reference architecture to support context-awareness feature, users security and privacy, while enabling the next generation networks integration. Figure 3 shows the integration of the proposed architecture into the ITU-T IoT reference architecture.\\
The management, control and data planes of the ITU-T IoT reference architecture need context-awareness capabilities to allow dynamic and flexible management of IoT networks (dynamic traffic steering, location-aware services, etc.). Therefore, the KP will be very useful for these planes of the ITU-T IoT reference architecture.
\begin{figure}
\includegraphics[width=\textwidth]{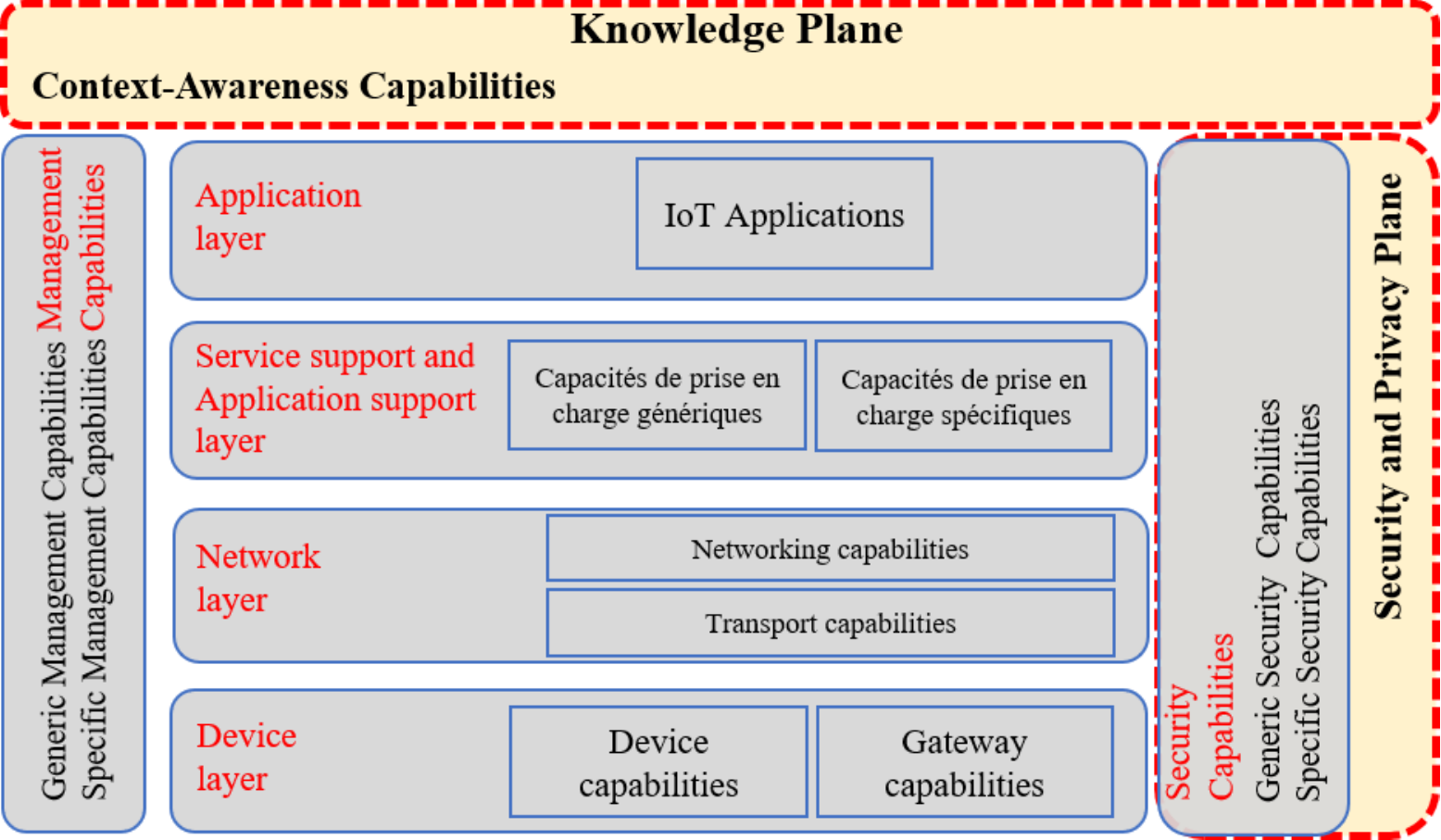}
\caption{ITU-T reference architecture integrating our proposed architecture} \label{appfig2}
\end{figure}\\
\newpage
\section{CASPaaS Underlying network architecture}
New network architectures pave the way in the development of service-oriented computing, enabling the deployment of ``as a service" architectures and virtualized environments in which only the necessary network function instances will be used when needed. They bring a new philosophy based on the transformations carried out in network architectures, essentially based on virtualization and network programming. They can thus support service-oriented computing, dynamic network programming through Software Defined Networking (SDN), Network Function Virtualization (NFV), Edge computing, etc.\\
\indent Based on these technologies, our architecture can be implemented as VNF (Virtual Network Function). Then, it can be deployed instantly in the network, regardless of the user's location. This will ensure an optimal security and privacy levels for the user wherever he is. Thanks to VNFs and service function chaining, it will be possible to dynamically orchestrate the deployment of the service as close as possible to the user \cite{vilalta_telcofog_2017}. Moreover, these new network paradigms fit to ITU-T IoT reference architecture. Indeed, their Management and Orchestration plane can extend the management layer of the ITU-T IoT reference architecture (Fig. 4).
\begin{figure}
\vspace*{-1cm}
	\includegraphics[width=0.96\textwidth]{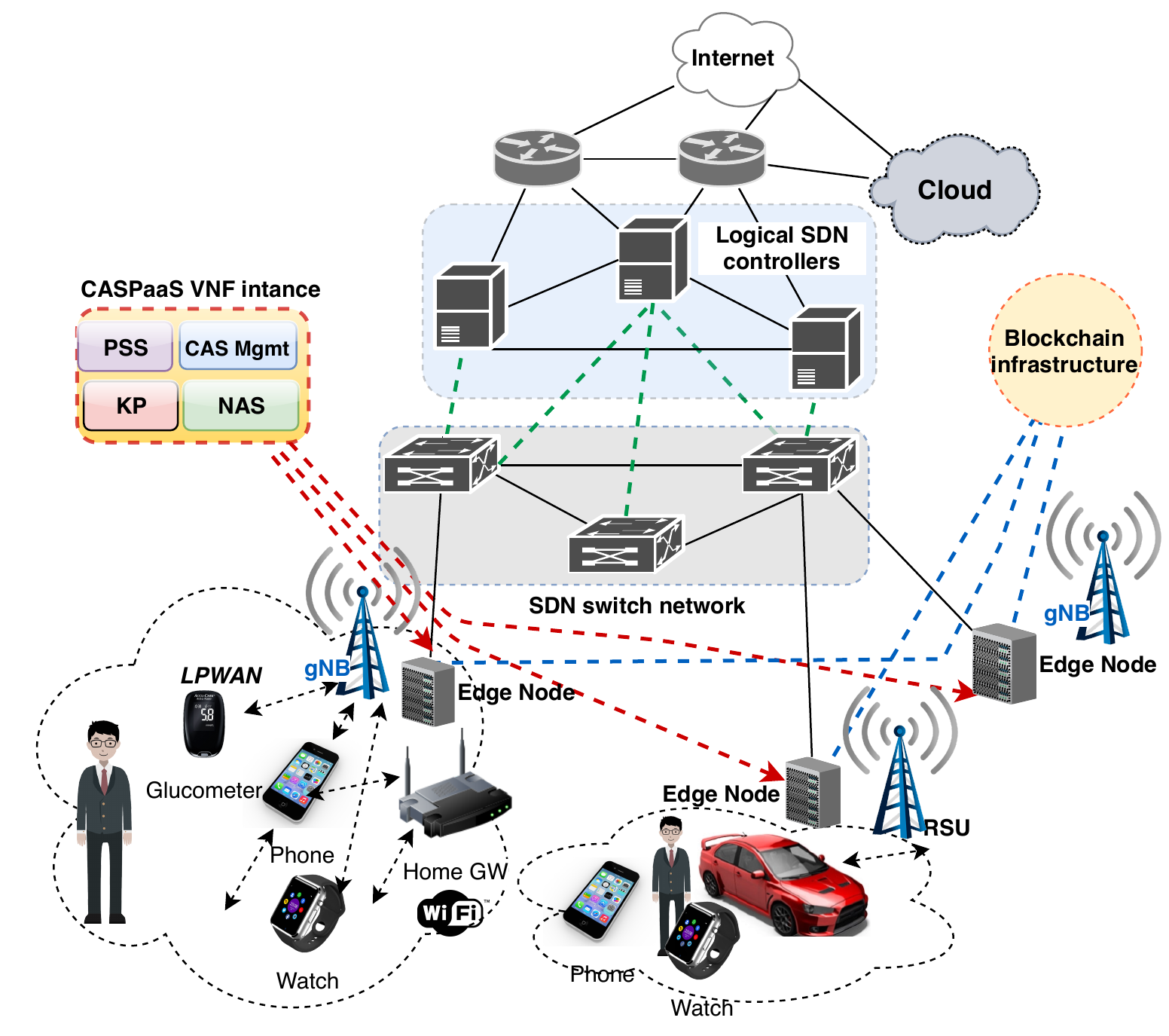}
	\caption{CASPaaS Architecture General View} \label{appfig6}
\end{figure}
\end{subappendices}
\end{document}